\documentclass[12pt,prd,showpacs,showkeys,preprintnumbers,amsmath,amssymb]{revtex4}

\usepackage{amsmath,amsfonts}

\newcommand{\BC}{{\mathbb C}}

\newcommand{\CO}{{\cal O}}

\newcommand{\ket}[1]{{|#1 \rangle }}

\newcommand{\tr}{\hbox{ Tr}}

\newcommand{\state}[1]{\mathopen{|}#1\mathclose{\rangle}}
\newcommand{\osc}[1]{\mathbf{#1}}

\def\[{\begin{equation}}
\def\]{\end{equation}}
\def\<{\begin{eqnarray}}
\def\>{\end{eqnarray}}

\newcommand{\nl}[1][0pt]{\nonumber\\[#1]&\hspace{-4\arraycolsep}&\mathord{}}

\newcommand{\earel}[1]{\mathrel{}&\hspace{-2\arraycolsep}#1\hspace{-2\arraycolsep}&\mathrel{}}
\newcommand{\eq}{\earel{=}}
\newcommand{\sfrac}[2]{{\textstyle\frac{#1}{#2}}}

\begin{document}
\title{Giant magnon bound states from strongly coupled ${\cal N}=4 $ SYM}
\author{David Berenstein$^{1}$,  Samuel E. Vazquez}
\email{dberens@physics.ucsb.edu}

 \affiliation{$^1$Department of
Physics, University of California
at Santa Barbara, CA 93106}

\begin{abstract}
We calculate in a very simple way
the spectrum of giant magnon bound states at strong coupling
in ${\cal N}=4$ SYM, by utilizing the description of the field theory vacuum 
in terms of a condensate of eigenvalues of commuting matrices.
We further show that these calculations can be understood in terms of
the central charge extension that permits the calculation of BPS masses in the Coulomb branch of ${\cal N}=4$ SYM. This paper shows further evidence that the strong coupling expansion
of the maximally supersymmetric Yang-Mills theory in four dimensions can be done systematically from first principles, without the assumption of integrability.

\end{abstract}
 \maketitle

\section{Introduction}

A program to study the strongly coupled ${\cal N}=4$ SYM theory from first principles was initiated in \cite{B}. Initially the proposal was thought to be a crude approximation
to the strong coupling limit of ${\cal N}= 4$ SYM theory. This approximation could heuristically account for
the sphere part of the geometry of the dual $AdS_5\times S^5$ string theory
\cite{M}, as well as
an estimate of the string tension. More evidence has shown that this approach may account systematically for the strong coupling expansion of ${\cal N}=4 $ SYM from first principles via purely field theoretical results \cite{BCV} (see also \cite{V}). In this paper we show that these techniques can be
applied to calculate the spectrum of bound states of giant magnons (these are described by classical solutions of the string in $AdS_5\times S^5$ \cite{HM}) from first principles without the assumptions of integrability \cite{Bei, Dorey}.

The approach to solve the theory at strong coupling can be motivated in the following way:
the action for the scalar fields of ${\cal N}=4$, when compactified on $S^3$, has a conformal coupling to
the curvature. Thus, we have that the Hamiltonian for the scalar fields is roughly
\begin{equation}
H \sim \tr \left[\int _{S^3} \frac12(\dot X)^2+\frac 12 X^2 +\frac {g_{YM}^2}4 [X,X]^2\right]\;,
\label{eq:hh}
\end{equation}
where the mass term is induced by such a conformal coupling to the curvature. Thus, at least at weak coupling, the theory is characterized by an oscillator description with no massless modes. The lightest modes are the the uniform field configurations, and if one wants to study systematically the lowest lying excitations, one should solve the problem for the lightest degrees of freedom first. It turns out that if one thinks of the constant modes of the sphere, one would typically expect that at weak coupling their wave function is characterized by a Gaussian matrix model. For such a matrix model, the typical eigenvalues for a hermitian matrix are of order $\sqrt N$.

We find this way that the kinetic term and the mass term is of order $(\sqrt N)^2 N\sim N^2$,
while the quartic potential is of order $g_{YM}^2(\sqrt N)^2 (\sqrt N)^2 N\sim \lambda N^2$, where $\lambda$ is the 't Hooft coupling constant $g_{YM}^2 N$ that controls perturbation theory \cite{'tH}. We see that as we take $\lambda$ to large values the potential term dominates the dynamics.

Thus one would expect that at strong coupling, and for small energies, we need to look at
the minima of the quartic potential. These occur for sets commuting matrices $[X^i,X^j]=0$.  These are hermitian matrices in the case of $U(N)$ or $SU(N)$ gauge group that we are studying.
Via a gauge transformation one can diagonalize the $X$ fields simultaneously, so the problem at strong coupling and low energies reduces to the dynamics of the eigenvalues of the $X$ matrices.

The kinetic and quadratic term should be thought of as a perturbation. Since the set of commuting matrices has a moduli space, we have an effective Hamiltonian dynamics on the set of constant commuting matrices, which is induced by the kinetic and quadratic terms.
One obtains this way a dynamics for $N$ identical bosons in six dimensions and there are a natural set of (essentially holomorphic) wavefunctions of these particles that one should consider \cite{B}. The eigenvalue dynamics contains enough information to reproduce the spectrum of
gravitons in the AdS/CFT: one shows that one obtains the same representations and counting of states as those at weak coupling \cite{GKP,Witten} (this is done by realizing that the highest weight states of the conformal multiplets for gravitons are all half-BPS states).

In the process of diagonalizing the matrices one finds measure factors that are very similar to a VanderMonde determinant. These measure terms produce an effective repulsion between the bosons in six dimensions, so the particles form a non-trivial geometric configuration in six dimensions, and the shape of this configuration depends on the exact details of the wavefunction. For the ground state, the shape of the particles is a hollow sphere \cite{B} whose radius can be calculated exactly \cite{BCV} , see also \cite{BCott} for numerical studies of these wavefunctions. One can also give a qualitative explanation of different spacetime topologies by comparing the intrinsic geometry of the eigenvalues with different gravity solutions \cite{LLM}.

In the work \cite{BCV} the study of the off-diagonal degrees of freedom was started. It was shown that, to leading order, the off-diagonal degrees of freedom accounted for the correct
dispersion relation of BMN defects to all orders in perturbation theory \cite{BMN}.
 This was
done by expanding about the correct geometry of the eigenvalues induced from the analysis of the wave function for the diagonal degrees of freedom.
Moreover, the correct dispersion relation for general magnons was recovered by these degrees of freedom, a result that was previously expected from \cite{SZ} and  was suggested to be correct by the expected integrability of the string in $AdS_5\times S^5$ \cite{MZ, BS,BDS}.
Another similar approach was pursued in \cite{Joao} with similar results.

Later on Beisert \cite{Bei} gave an argument for the exactness of this dispersion relation in terms of BPS multiplets of
a central extension of the set of charges on an infinite spin chain associated to ${\cal N}=4$ SYM. The central charge is only allowed for infinite strings. For finite strings it vanishes, basically due to the Virasoro constraints (these are realized by the cyclic property of the trace \cite{BMN}).

Moreover the field theory analysis suggested a geometric interpretation of the magnon quasimomentum $p$ (the phases between different terms in a typical BMN state)  on the worldsheet as a geometric angle in the $S^5$ geometry. This was later verified by  Hofman and Maldacena \cite{HM} by studying the classical string motion in $AdS_5\times S^5$, where they showed that some of these solutions associated to large angles corresponded to the correct dispersion relation. These were dubbed giant magnons. Dorey argued that bound states of giant magnons are also BPS with respect to this central charge extension \cite{Dorey}, and this permits an evaluation of the dispersion relation of these bound states.

In this paper we will find that this dispersion relation calculated by Dorey can also be obtained using the
techniques of \cite{B,BCV}. We will also explain why the result can be naturally associated to the computation of central charges on the Coulomb branch of ${\cal N}=4 $ SYM, providing
a raison d'etre for the BPS nature of the calculation, giving us some field theoretical interpretation of the worldsheet central charge of Beisert.

The computation of the dispersion relation for these bound states via matrix model methods has also been claimed by the paper \cite{HO}, but there are some technicalities in the procedure to arrive at the result that do not apply at strong coupling. We will describe these in detail  later on and we will explain how our computation differs from this one.

It has also been found that these ideas also work for $q$
deformations of ${\cal N}=4 $ SYM, \cite{BC}, abelian
orbifolds of ${\cal N}=4$ SYM \cite{BCottO}, and the computation of string energies seems to work for other bubbling geometries as well \cite{CCC}.
Similar techniques have also been required to solve aspects of the ${\cal N}=4$ SYM field theory at finite temperature \cite{GHHK,AH}. In this sense, the
strong coupling analysis satisfies most of the tests of the AdS/CFT correspondence for ${\cal N}=4$ SYM that have been performed at weak coupling, with the bonus that at at strong coupling we naturally see geometrical concepts arise from the dynamics of the field theory. In this sense, we have emergent geometry from strong coupling dynamics. This paper puts more  evidence in favor of this strong coupling expansion.

\section{Computation of the off-diagonal oscilator energies}

We want to do a calculation of energies of dual strings in ${\cal N}=4$ SYM theory at strong coupling from first principles. We will follow closely the techniques developed in \cite{B,BCV}.
These will not be reviewed in detail, and the reader is urged to consult these works as the need arises.
The basic procedure starts by first identifying a subset
of the degrees of freedom that consist of
constant scalar matrices in ${\cal N}=4$ SYM.

The effective (truncated) Hamiltonian for these degrees of freedom is
\begin{equation}
H \sim \tr \left[ \sum_{i=1}^6 \frac12(\dot  X^i )^2+\frac 12 ( X^i)^2 +\sum_{i,j<6} \frac  {g_{YM}^2}{8\pi^2 }[X^i,X^j]^2\right]\;.
\label{eq:setp1}
\end{equation}
This follows from equation (\ref{eq:hh}) by integrating over the volume of the sphere and by normalizing the matrix degrees of freedom so that they have canonical form. The factors of
$\pi$ are induced by the volume of the sphere.

The premise of the works \cite{B,BCV} is that at strong coupling the degrees of freedom of ${\cal N}=4$ SYM theory are organized differently than at weak coupling.
In particular, it was argued that at low energies and at strong coupling only configurations of commuting matrices
dominate. After using the gauge invariance of the model, one can turn the problem into eigenvalues of the six matrices. The six eigenvalues associated to each diagonal entry of the matrices, $X_{ii}$, can be collected into a collection of six vectors $\vec x_{ii}$.
 The vacuum for the eigenvalues is characterized by a non-trivial
distribution density of eigenvalues $\rho(\vec x)$ (we will call these the diagonal condensate) where all eigenvalues are at the same distance from the origin $r_0$, forming a five sphere. The density in the large $N$ limit is given by
\begin{equation}
\rho(\vec x) \sim \delta(|\vec x|-r_0)\;,
\end{equation}
where $\int \rho(\vec x)= N$ and $r_0= \sqrt{N/2}$ (we are following the conventions of \cite{BCV}. This has also been verified numerically \cite{BCott}.

This distribution of eigenvalues should be treated as a non-perturbative
background field for the off-diagonal modes, that are treated perturbatively. This is, we ignore the back-reaction of the diagonal modes to the presence of the off-diagonal modes.

In this way one finds that the energies of the off-diagonal modes (treated quadratically) are given by \cite{BCV}
\begin{equation}
E_{jj'} = \sqrt{1+\frac {g_{YM}^2}{2\pi^2} |\vec x_j-\vec x_{j'}|^2}\;.
\end{equation}

The first term inside the square root is from the standard mass term $\vec X^2$, induced from the curvature coupling to the metric, while the second term is induced from the commutator squared, that has been expanded to quadratic order in the off-diagonal modes. Written in this form the Hamiltonian on the $S^3$ is normalized exactly so that it measures the dimension of operators under the operator-state correspondence for a Conformal Field Theory.
The mass equal to one corresponds to an operator with dimension equal to one. Thus the factor of $1$ inside the square root is the classical dimension of the field $X$ itself.

The idea for  this paper is to include the off-diagonal degrees of freedom for the whole tower of Kaluza-Klein modes on the sphere and to compute their energies, under the assumption that
the eigenvalues do not backreact (the are treated as a fixed background). To do this, we decompose the field $X$ into scalar spherical harmonics on the sphere. These are characterized by representations of the isometry group
of the $S^3$ sphere: $SO(4)\sim SU(2)\times SU(2)$. The representations that appear are
in the $(n/2,n/2)$ representation of $SU(2)\times SU(2)$. Let us call these modes $X_{(n)}$.

The energy for these modes contains three pieces. There is a commutator term, just like in
the case of the $n=0$ states. There is also the curvature coupling, which contributes $1$ to the mass, and there is the gradient term on the sphere.

The gradient terms and the curvature coupling are of the form
\begin{equation}
\int_{S^3} |\nabla X_{(n)}^2|+ |X_{(n)}|^2 \sim m_{(n)}^2 |X_{(n)}|^2\;,
\end{equation}
and give an effective mass for each of these spherical harmonic modes, even in the free field theory. The off-diagonal modes are complex $X_{jj'}$ and $ X_{j'j} =X^*_{jj'}$.

In the free field theory one can relate these modes to the list of operators via the operator state correspondence, and one finds that the modes are related by
$X_{(n)} \sim \partial^{(n)} X$ (more precisely, it is the raising mode for the corresponding oscillator that is mapped to the operator). Because the
operator $ \partial ^{(n)} X$ has dimension $(n+1)$, this should be the energy of the corresponding oscillator mode on the three sphere. Thus we find the simple result that
\begin{equation}
m^2_{(n)}|_{free} = (n+1)^2\;,
\end{equation}
so that the energy of the oscillator and the dimension of the operator agree.
This is a quick way to get at the final answer.
This can also be evaluated explicitly using spherical harmonics on the sphere.

At strong coupling, we also get an extra term from the commutator that includes the diagonal condensate information. This term is identical to the one for the $n=0$ modes. Thus we find that the full masses of the off-diagonal tower of degrees of freedom is given by
\begin{equation}
E_{jj'(n)} = \sqrt{(n+1)^2 + \frac {g_{YM}^2}{2\pi^2} |\vec x_j-\vec x_{j'}|^2}\label{eq:kkod}\;.
\end{equation}

 The calculation above requires us to be careful with gauge invariance.
The scalar modes that have these masses are such that their polarization satisfies
$\vec \delta X_{jj' (n)} \cdot (\vec x_j -\vec x_{j'}
) = 0$. The other (sixth) polarization is sensitive to non-constant gauge transformations, and necessarily
mixes with the off-diagonal $W$ bosons, providing the extra polarizations required by the Higgs mechanism. A similar calculation can be done with the off-diagonal modes for the fermions and the vector fields, but it will not be done here.

\section{Chain localization argument and strong coupling}

To understand how the calculation done in the previous section relates to
the spectrum of giant magnons, we need to review some of the results found in \cite{BCV}.
The idea is to translate BMN type operators to states in the quantum dynamics
on the sphere. This is easy to do at weak coupling. However, at strong coupling we
need to consider states that naturally mimic the physics of the weak coupling spin chain and the
matching of BPS states with strong coupling states.

Let us consider operators that in the free field theory are of the form
\begin{equation}
\CO_k \sim \sum_{j}  \tr( \dots Z^j Y Z^{l-j}\dots) \exp(i k j)
\end{equation}
as in a typical BMN state \cite{BMN}. These can also be written for $k\neq 0$ as
\begin{equation}
\CO_k \sim \frac 1{1-\exp(ik)} \sum_{j}  \tr( \dots Z^j [Y,Z] Z^{l-j-1}\dots) \exp(i k j)\;.
\end{equation}

This second form of describing the operator makes it obvious that at least $Y$ can not be considered to commute with $Z$, as otherwise the operator would vanish.

Without the $Y$ impurity, the $Z$ background
\begin{equation}
\CO_k \sim \frac 1{1-\exp(ik)} \sum_{j}  \tr( \dots  Z^{l}\dots)
\end{equation}
is to be treated as BPS. At strong coupling, the BPS excitations are just eigenvalue wavefunctions \cite{B}. Thus, it is consistent to treat the $Z$ background as a diagonal object
in the dual state to the operator $\CO_k$. Thus $Y$ can be thought of as
being a particular off-diagonal excitation connecting two eigenvalues $m,m'$, where
$Z\sim \hbox{diag}( \dots z_m \dots z_m')$.

The corresponding state associated to the operator $\CO_K$ is given by
\begin{equation}
\ket{ {\CO_k } } \sim \sum_{m m' j}  \tr( \dots z^j_m Y^\dagger_{m m'} z_{m'}^{l-j}\dots) \exp(i k j) \label{eq:odd}\;,
\end{equation}
where we have turned on a particular off-diagonal excitation. However, we see that we are summing over $m,m'$, namely, the distribution of eigenvalues. This sum is necessary to make the state gauge invariant: the permutation of eigenvalues is a residual gauge symmetry of diagonal matrices that makes the eigenvalues behave as identical particles. Summing over eigenvalues implements this gauge invariance constraint.

If we consider the equation (\ref{eq:odd}) as a ket for the off-diagonal modes in a fixed background distribution of the eigenvalues, we find that the wave function is a superposition of different off-diagonal modes with different amplitudes, and that these amplitudes are proportional to large powers of the eigenvalues $z_m$, $z_m'$. Thus, in the above expression the oscillators that contribute the most to this ket are those for which $|z_m|$ and $|z_m'|$ have maximum value.

The distribution of the eigenvalues is a five-sphere. The quantities $|z_m|$ is the radius of the projection of the five sphere onto a two dimensional disc (along the 12 plane let us say).
It is easy to see that the eigenvalues that dominate are on the edge of the disc. Both of these correspond to a particular diameter of the five sphere along the $12$ plane.

Thus, we have that $z_m \sim r_0 \exp(i\phi_m)$, and $z_m'\sim r_0 \exp(i\phi_m')$ where $\phi_m$ and $\phi_m'$ are phases, and $r_0$ is the radius of the sphere.

Having fixed $m,m'$, we also have a sum over phases, that is of the form
\begin{equation}
\sum_{j} \exp( i j \phi_m  - i j \phi_{m'} + i j k )\sim \frac {1-\exp(i (\phi_m-\phi_{m'}+k)j)}
{1-\exp(i (\phi_m-\phi_{m'}+k))}\;.
\end{equation}
This sum has a large value when $\phi_m-\phi_m'+k \sim 0$, that is, when all the
summands have the same phase. In that case the sum is of order $j$, while for other values
of the phases it mostly cancels. As probabilities are controlled by the amplitude squared, we have that the ket is dominated by eigenvalues $z_m, z_m'$ whose phase differs exactly
by $k$. This is, $k$ is represented geometrically by an angle in the $12$ plane.

This argument states that the dominant contribution to the energy of this state arises from localization onto particular pairs of eigenvalues whose locations can be read from the parameters describing the state.

The energy of the off-diagonal degree of freedom $Y$ for this state will be
\begin{equation}
E\sim \sqrt {1+\frac{g_{YM}^2 r_0^2}{2\pi^2} \sin^2( k/2)}\;.
\end{equation}
This is the result that was reported on \cite{BCV} and it reproduces the giant magnon dispersion relation (see \cite{BDS, SZ}). This prediction of the angle was later verified geometrically in \cite{HM}.

Now, we can do the same calculation with the other off-diagonal oscillators corresponding
to the Kaluza Klein tower of scalars compactified on the sphere. We use the same localization argument as above, but we insert different off-diagonal KK modes in the description of the state. This is, we replace $Y^\dagger_{m,m'}$ with $Y^\dagger_{(n), m,m'}$.

In the free field theory limit, we would be looking at operators of the form
\begin{equation}
\CO_{n,k} \sim \sum_{j}  \tr( \dots Z^j (D^{n} Y) Z^{l-j}\dots) \exp(i k j)\;.
\end{equation}
This is using the fact that the higher spherical harmonics of $Y$ on the sphere are related via the operator state correspondence to $\partial^{(n)}Y$. This has to be covariantized to make a gauge invariant operator. This is why we have $D^{n} Y$ instead of $\partial^{(n)} Y$.

If we follow the same localization reasoning as before, we would get an energy of the form
\begin{equation}
E\sim \sqrt {(n+1)^2+\frac{g_{YM}^2 r_0^2}{2\pi^2} \sin^2( k/2)}\label{eq:gmbs}\;,
\end{equation}
where now we calculate the energy for the corresponding oscillator.

This result is numerically identical to the dispersion relation for the BPS bound state of $n+1$ giant magnons
described by Dorey \cite{Dorey}. However, in the case described by Dorey, it was a bound state of $(n+1)$ Y defects, whereas here it describes a single off-diagonal oscillator.

If we go to weak coupling, we know that the operator
\begin{equation}
\CO_{n,k} \sim \sum_{j}  \tr( \dots Z^j (D^{n} Y) Z^{l-j}\dots) \exp(i k j) \label{eq:op}
\end{equation}
mixes with
\begin{equation}
\CO_{n,k} \sim \sum_{j}  \tr( \dots Z^{j_1} D Z \dots DZ \dots Z  Y Z
Z\dots DZ\dots) \exp(i k j)\;,
\end{equation}
so it can be thought of as a bound state of $(n)$ $D$ type impurities with one
$Y$ impurity. This problem usually leads to a spin chain that can be solved via a
Bethe Ansatz \cite{BS}.

On a bound state, in the Bethe ansatz spin chain model, one is at a zero or pole of the scattering matrix and this happens for complex values of the quasimomentum. In these bound states, the
amplitude for the operator where the derivative is at distance $l$ from $Y$
\begin{equation}
O_{n,k,l}\sim \tr ( Z\dots Z Y Z^l DZ \dots)\;,
\end{equation}
decays exponentially relative to the situation where the $D$ is on top of the $Y$, by the imaginary part of the quasimomentum.


At weak coupling, one can see the presence of this bound state by directly diagonalizing the one-loop dilatation operator in the $SU(1,1|2)$ subsector. This was done in \cite{BS2}, and we will now quote the main results. We take the vacuum to be $| 0 \rangle = |Z Z \cdots Z \rangle$. The elementary excitations at each site are given by
\[ Z\rightarrow D Z\;,\;\;\; Z\rightarrow Y\;,\;\;\; Z\rightarrow \cal U\;,\;\;\; Z\rightarrow \dot{\cal U}\;,\]
where $\cal U$ is the gaugino and $\dot{ \cal U}$ is an additional fermion. Note than a transition like,
$Z \rightarrow DY$, can be interpreted as a double excitation.

One can now introduce a convenient set of bosonic and fermionic oscillators at each site corresponding to elementary excitations, where the standard excitations above are composite 
objects. More specifically, we write
\[
\state{Y}=\osc{c}^\dagger\osc{\dot c}^\dagger\state{Z},\qquad
\state{{\cal U}}=\osc{a}^\dagger\osc{\dot c}^\dagger\state{ Z},\qquad
\state{\dot{{\cal U}}}=\osc{c}^\dagger\osc{\dot a}^\dagger\state{Z },\qquad
\state{D Z}=\osc{a}^\dagger\osc{\dot a}^\dagger\state{ Z},
\]
where $\osc{a}, \osc{\dot a}$ are bosonic oscillators, and  $\osc{c}, \osc{\dot c}$ are fermionic oscillators.

In \cite{BS2}, the two-impurity energy eigenstate  was found to be of the form
\<\label{twoimp}
\state{p_{\osc{A},\osc{\dot A}};q_{\osc{B},\osc{\dot B}}}
\eq
\sum_{\ell_1<\ell_2}e^{ip\ell_1+iq\ell_2}
\osc{A}^\dagger_{\ell_1}
\osc{\dot A}^\dagger_{\ell_1}
\osc{B}^\dagger_{\ell_2}
\osc{\dot B}^\dagger_{\ell_2}\state{0}
\nl
+\sum_{\ell_1=\ell_2}
\frac{u-v}{u-v-i}\,
e^{ip\ell_1+iq\ell_2}
\osc{A}^\dagger_{\ell_1}
\osc{\dot A}^\dagger_{\ell_1}
\osc{B}^\dagger_{\ell_1}
\osc{\dot B}^\dagger_{\ell_1}\state{0}
\nl
+\sum_{\ell_1>\ell_2}
\frac{(u-v)^2}{(u-v-i)(u-v+i)}\,
e^{ip\ell_1+iq\ell_2}
\osc{A}^\dagger_{\ell_1}
\osc{\dot A}^\dagger_{\ell_1}
\osc{B}^\dagger_{\ell_2}
\osc{\dot B}^\dagger_{\ell_2}\state{0}
\nl
+\sum_{\ell_1>\ell_2}
\frac{i(u-v)}{(u-v-i)(u-v+i)}\,
e^{ip\ell_1+iq\ell_2}
\osc{A}^\dagger_{\ell_2}
\osc{\dot A}^\dagger_{\ell_1}
\osc{B}^\dagger_{\ell_1}
\osc{\dot B}^\dagger_{\ell_2}\state{0}
\nl
+\sum_{\ell_1>\ell_2}
\frac{i(u-v)}{(u-v-i)(u-v+i)}\,
e^{ip\ell_1+iq\ell_2}
\osc{A}^\dagger_{\ell_1}
\osc{\dot A}^\dagger_{\ell_2}
\osc{B}^\dagger_{\ell_2}
\osc{\dot B}^\dagger_{\ell_1}\state{0}
\nl
+\sum_{\ell_1>\ell_2}
\frac{i^2}{(u-v-i)(u-v+i)}\,
e^{ip\ell_1+iq\ell_2}
\osc{A}^\dagger_{\ell_2}
\osc{\dot A}^\dagger_{\ell_2}
\osc{B}^\dagger_{\ell_1}
\osc{\dot B}^\dagger_{\ell_1}\state{0}.
\>
The rapidities $u,v$ are defined via
\[
e^{ip}=\frac{u+\sfrac{i}{2}}{u-\sfrac{i}{2}}\,,\qquad
e^{iq}=\frac{v+\sfrac{i}{2}}{v-\sfrac{i}{2}}\,.
\]
The first line in (\ref{twoimp})  represents the incoming excitations.
The second line represents the wave-function
when the two excitations overlap, and the remaining four lines represent
outgoing excitations and they encode the S-matrix.

We can now consider an incoming $Y$ and $DZ$ excitations. This is done by setting $\osc{A} = \osc{a}$, $\osc{{\dot A}} = \osc{{\dot a}}$, $\osc{B} = \osc{c}$ and $\osc{{\dot B}} = \osc{{\dot c}}$.  Looking at the outgoing states in (\ref{twoimp}) we see that there is a bound state for quasi-momenta satisfying
\[ \label{pole} u - v - i = 0\;.\] Following \cite{Dorey}, we can solve this condition by setting $p  = P/2 + i s$, $q = P/2 - i s$, where $P$ is the total momentum of the impurities. One then quickly finds that (\ref{pole}) is satisfied provided $s = \log \cos(P/2)$. The total energy of the excitations is then given by \cite{BS2},
\[ E = {\cal E}(p) +  {\cal E}(q)  =  \frac{g_{YM}^2 N}{8\pi^2}\left(4\sin^2 (p/2) +  4\sin^2 (q/2) \right)=  \frac{ g_{YM}^2 N}{4 \pi^2} \sin^2(P/2)\;.\]
This agrees precisely with  the one-loop expansion of the anomalous part  of (\ref{eq:gmbs}) for $n = 1$.

Looking back at the state (\ref{twoimp}), we see that the amplitudes for the outgoing states are indeed exponentially suppressed  as $\sim \exp( i p l_1 + i q l_2)  = e^{i P} \exp[- (l_1 - l_2) |\log \cos (P/2)|]$, where $l_1 \geq l_2$. Therefore, the bound state wavefunction is dominated by the state where the two excitations are in top of each other (the second line of (\ref{twoimp})) which, in our case, is the double excitation $Z \rightarrow D Y$.
The strong coupling calculation we have done seems to indicate that as we increase the coupling this state dominates more and more and provides the full answer for the energy.

Since the bound state we are calculating has the same dispersion relation as the bound state
of $(n+1)$ Y magnons, we would like to conjecture that they belong to the same multiplet of the spin chain. It is known that these magnon bound states of the $PSU(1,1|2)$ sector are highly degenerate \cite{BS2}. This has recently been explained in \cite{BZ}. The best way to show this is to provide the symmetry that transforms one state into the other one. We have not been able to show this directly. Instead, we will give an indirect argument.

Before we do that, we would like to compare our results with those that were reported in \cite{HO}.
The calculation found in \cite{HO} considers a bound state of $(n+1)$ Y impurities.
They argue that in the case of the $n+1$ impurities, the off-diagonal modes will add
to the energy as
\begin{equation}
E_{total} = \sum \sqrt{1+\frac {g_{YM}^2}{2\pi^2}|\vec x_s - \vec x_{s+1}^2|}\;,
\end{equation}
and that this off-diagonal energy is minimized (for fixed endpoints of the first and last $Y$, $\vec x_0$, and $\vec x_{n+1}$)
by having all $\vec x_s-\vec x_{s+1}$ being equal to each other, and equal to
\begin{equation}
\frac 1{n+1}(\vec x_0-\vec x_{n+1})\;.
\end{equation}

At strong coupling, the distribution of eigenvalues forms a sphere that is empty, so there are no eigenvalues at the required positions to make the argument work in a naive way.
In the plane wave limit the argument is correct, as that is the setup where the segment representing the bound state is essentially tangent to the sphere, so the required eigenvalues are available.

At weak coupling, the important eigenvalue distribution is the one for many $Z$, ignoring the
other degrees of freedom, and this is a disc \cite{Btoy}. In this sense, the calculation \cite{HO} is applicable at weak coupling, and more closely resembles the approach to calculate BMN energies found in \cite{Joao}, where one studies a two matrix model by first solving exactly the one matrix model, and using this information to perturb the energies of the oscillators of the other matrix.
We believe that it is  the process of minimization of the energy that is responsible for giving the correct answer for the bound state. The reason for this is that it is believed that these bound states are BPS \cite{Bei, Dorey} along the spin chain,
so in the process of minimizing the energy one is moving towards saturating a BPS bound.

We now want to argue that the result we found can also be massaged into a way that
associates it closely to BPS states in the ${\cal N}=4 $ SYM theory and therefore is
exact. We want to claim that it is this connection with a BPS bound that is saturated that tells us that the
state belongs to the giant magnon bound state multiplet.

\section{Interpretation of the computation as a central charge}

Let us consider an ${\cal N}=2$ SYM theory in four dimensions, with some matter content.
These theories usually have a moduli space of vacua characterized by the vacuum expectation values of the scalar superpartner of the gaugino. The unbroken gauge symmetry of
these configurations is $U(1)^r$, where $r$ is the rank of the gauge group.
Associated to each point in the moduli space, there is a spectrum of massive $W$ bosons and
also non-perturbative excitations (monopoles and dyons). The massive $W$ bosons
have fewer polarizations than a generic representation of the ${\cal N}=2$ superalgebra, and
therefore they are in BPS multiplets that saturate a BPS bound
\begin{equation}
M(W) = |Z(W)|\;,
\end{equation}
where $Z(W)$ represents the (complex) central charge extension of the ${\cal N}=2 $ Poincare superalgebra. Understanding the properties of the spectrum of the central charge $Z$ and of BPS states can be used to solve the theory in the infrared \cite{SW}. Generically the central charge is described by a lattice of charges, and the lattice is related to the root lattice of the gauge group algebra. Thus, the central charge is correlated with the charges of objects under the unbroken $U(1)^r$ gauge symmetry  (the electric and magentic charges of the corresponding particle). This was explained in detail in \cite{WO}.

The superpartner of the vector, $\phi$, has a vev in the Cartan, and the classical value of the central charge for an electric particle of weight $w$ is given by $g_{YM} w \cdot \phi$, thus it is a linear combination of the vevs of $\phi$.

For the case of $SU(n)$, these formulas give $Z_{jj'} \sim g_{YM} (\phi_j-\phi_j')$ for an off diagonal $W$ boson connecting the $j, j'$ eigenvalues of $\phi$.
For a monopole the mass would be of order $ g^{-1}_{YM} (\phi_j-\phi_j')\sim Z_{el}/g_{YM}^2$.
These formulas generally receive loop corrections from wave function renormalization and non-perturbative corrections \cite{SW}.

For the case of ${\cal N}=4$ SYM the classical result for the masses is exact. However, one generally has to choose an ${\cal N}=2 $ superalgebra with respect to which the corresponding states are BPS. The central charge is not anymore a single complex number, but it is a list of various numbers, plus a choice of orientation for the charge. This is also correlated with the fact that the moduli space of vacua for $U(N)$ SYM corresponds to $(\BC^3)^N/S_N$, and
the mass of the off-diagonal modes are
\begin{equation}
m_{jj'}\sim g_{YM} |\vec\phi_j -\vec \phi_j'|\;.
\end{equation}

The simplest way to understand the central charge extension of ${\cal N}=4 $ SYM is to realize the system on a collection of parallel branes in type IIB string theory. The supersymmetry algebra in flat ten dimensions is described by 16 left mover supersymmetries and sixteen right moving supersymmetries on the string worldsheet.

The total momentum of a closed string is the sum of the left and right-moving momenta, namely
\begin{equation}
P^I\sim \oint \partial X^I+\bar \partial X^I\;,
\end{equation}
and the central charge ends up being the difference of the two
\begin{equation}
Z^I\sim \oint \partial X^I-\bar \partial X^I\;,
\end{equation}
which is associated with winding of the string on toroidal compactifications.

For open strings suspended between D-branes, one has a partial winding in the directions that
are transverse to the D-brane. Thus the central charge of ${\cal N}=4 $ SYM measures exactly this partial winding, and the mass of the lightest string is just the distance between the D-branes \cite{DLP}.

If we take a massive such particle, and we boost it, we obtain a dispersion relation of the form
\begin{equation}
E \sim \sqrt {p^2 + (Td_{jj'})^2}\label{eq:odW}\;,
\end{equation}
where $T$ is a properly normalized string tension, and $d_{jj'}$ is the distance between the D-branes. It is well known that the eigenvalues of the scalar fields in ${\cal N}=4 $ SYM
represent the position of the branes  (this was instrumental in stating that the dimensional reduction of ten dimensional SYM to quantum mechanics was describing M-theory in the lightcone \cite{BFSS}).

We want to use this same intuition for the case of
${\cal N}=4 $ SYM compactified on the sphere. As has been described previously,
the vacuum is characterized by a non-trivial distribution of eigenvalues.

If we were on flat space, the fact that the eigenvalues are separated makes it possible
for us to use the intuition about the central charge extension described above. However, in the case of the sphere, gauge invariance makes it impossible to have just a single off-diagonal mode excited.

What we can do instead is consider the problem from the point of view of energy scales. At strong coupling, the off-diagonal modes are heavy, and their typical mass is of order $\sqrt \lambda$ relative to the radius of the sphere. Thus, the Compton wavelength for the off-diagonal modes is much shorter than the radius of the sphere, and this is the typical localization length of these particles on the sphere.
In this sense, the off-diagonal excitations (the heavy W boson multiplets) are not particularly sensitive to the radius of the sphere, except as a global quantization condition of their momenta, and can be treated as if they were in flat space.

Since we have at least two of these off-diagonal modes, we want to make sure that they are well separated and not forming a bound state, so that we can think of them as a collection of separated BPS particles
in flat space whose energy can be computed exactly.
This can be done by storing energy on the branes as photons (or their superpartners) so that the particles are dissociated by the photons.

This is what the background of $Z$ excitations is providing for us: it is exciting the degrees of
freedom of the branes that the off-diagonal modes are touching. In this approximation, the
energy of the system is a sum over the different BPS particles plus the photons,
and for each of these $W$ bosons the energies are of the form
\begin{equation}
E_i \sim \sqrt{ p^2 + m_i^2}\;,
\end{equation}
where $m_i$ is the mass of the corresponding off-diagonal W-boson. This is the same as
what appears in equation (\ref{eq:odW}), but $p^2$ needs to be interpreted as the momentum on the sphere. Since the scalar modes of the $W$ boson are conformally coupled to the
background metric, the eigenvalues for $p^2$ that we get are exactly $(n+1)^2$ in untis where the radius of the sphere is equal to one.
The square root formula is interpreted as the standard relativisitic dispersion relation for a particle in flat space, and the mass can be interpreted as the mass of a BPS excitation.

As a matter of fact, the whole square root formula saturates a flat space BPS bound where we can treat the momentum as a central charge of the algebra (this is exactly what one does if one thinks of the problem in dimensional reduction, or under T-duality, where momentum becomes winding in the T-dual direction). Thus the square root formula should be exact.

The main assumption that is made for the calculation is that the off-diagonal modes do not cause backreaction on the diagonal degrees of freedom. This requires that the vevs of the diagonal degrees of freedom should be considered to be large in natural $\hbar$ units, so that the presence of the off-diagonal excitations does not affect their average values substantially.
As the eigenvalues are of order $\sqrt N$, they are large only in the large $N$ limit. Thus the exactness of the mass formula depends crucially on being at large $N$.

We conclude that the dispersion relation formula we have found  in the previous section, as described in (\ref{eq:gmbs}), can be related to the central charge extension of ${\cal N}=4 $ SYM on the moduli space of vacua and therefore should be exact.

This fits very nicely with the observation of Beisert \cite{Bei}|
that the magnon dispersion relation is exactly captured by a central extension appearing
in the spin chain algebra at infinite spin chain length. This central charge extension has been
explored also in \cite{AFPZ}, where it was shown to arise by relaxing the level matching constraints. As a matter of principle, this is exactly what one does in toroidal compactifications of strong theory, where the level matching constraint receives contributions from winding
the string on tori, so the left moving level is different than the right moving level. This mismatch accounts for the central charge.
This same winding charge is associated with the masses of open strings stretching
between separated branes (as we have discussed above) .

Notice that the presence of a central charge extension in our calculation does not depend on the string being integrable, but it is instead a property of the ${\cal N}=4 $ SYM in
flat space. Indeed, in \cite{HM} it was argued that in the large $J$ limit the
ends of the giant magnon act as if they are attached to branes. Here we see that
they are attached to specific eigenvalues, and it is well known that eigenvalues have interpretations of branes in this context as well \cite{HHI}.

Under the assumption that the string worldsheet is just a mirror of what happens in spacetime, we find a new natural justification for the presence of a central charge extension in
the infinite spin chain limit of the $PSU(2,2|4)$ algebra.

\section{Conclusion}

In this paper we have seen that it is possible to do a  first principles calculation at strong coupling  in ${\cal N}=4$ SYM theory that
matches the dispersion relations of bound states of giant magnons \cite{Dorey}.
We have also seen that our calculations explain the origin of this dispersion relation as a BPS
quantity, because we can relate it to the spectrum of W-bosons in the Coulomb branch of ${\cal N}=4$ SYM. These $W$ bosons are BPS and their mass is calculated by a central charge extension of ${\cal N}=4 $SYM on flat space. This explanation does not require integrability, but it gives a different reason for the central charge extension of the string symmetry algebra \cite{Bei}.

From the point of view of integrable models, the fact that we get the correct energy and dispersion relation for the bound states indicates that the methods that are being explored here are compatible with a integrable structure of the string worldsheet. In particular, the energy and momentum of the bound state can be determined by the poles of the $S$-matrix scattering of magnons. Since we find a result in agreement with the expectations from integrability, it means that the strong coupliing expansion contains the information of the location of the poles in the worldsheet scattering matrix.

It would be interesting if one could also compute the dressing phase of the scattering matrix, but this seems to be prohibitively hard, as one would want to compare it with the results based on integrability \cite{BHL}.

It would also be interesting to see if it possible to calculate $1/J$ corrections to energies of states and see if one is able to match the string computations ( see \cite{RTT} and references therein), as well as the energies of leading twist operators \cite{GKP2}, whose anomalous dimension grows as the logarithm of the spin.

Overall, we feel that this approach to study strong coupling physics  for ${\cal N}=4 $ SYM has shown to be reliable and it matches a lot of physics that is known from other methods.  Although the systematics of this expansion still need to be explored further, so far the results are very promising and these expansions deserve to be studied further.

\section*{Acknowledgements}

D.B. would like to thank N. Dorey, S. Hartnoll for various discussions and correspondence related to this project. Work supported in part by  DOE, under grant DE-FG02-91ER40618.

\end{document}